\documentclass[aps,prd,preprint,nofootinbib]{revtex4-1}
\usepackage{amsfonts}
\usepackage{mathrsfs}
\usepackage{graphicx}
\usepackage{amsmath}
\usepackage{amssymb}
\usepackage{epsfig}
\usepackage{graphicx}
\usepackage{slashed}
\usepackage{color}
\usepackage{multirow}
\usepackage{tabularx} 
\usepackage{braket}
\usepackage{bm}
\usepackage{}
\newcommand{\appropto}{\mathrel{\vcenter{\offinterlineskip\halign{\hfil$##$\cr
				\propto\cr\noalign{\kern2pt}\sim\cr\noalign{\kern-2pt}}}}}

\usepackage{tikz} 

\usepackage[compat=1.0.0]{tikz-feynman}

\usepackage{ulem}
\usepackage{adjustbox}

\usepackage{stmaryrd}
\newcommand{\bqa}{\begin{eqnarray}}
	\newcommand{\eqa}{\end{eqnarray}}
\newcommand{\beq}{\begin{equation}}
	\newcommand{\eeq}{\end{equation}}
\allowdisplaybreaks[1]

\graphicspath{{fig/}{dia/}} \DeclareGraphicsExtensions{.eps}
\begin{document} 
\title{
Large CP Asymmetries from Final-State Interactions in Charmful Baryonic Decays of $B^0 \to \Xi_c^+ \overline{\Xi}_c^-$  and  $B_s^0 \to \Lambda_c^+ \overline{\Lambda}_c^-$
}
\author{Chao-Qiang Geng$^1$, Xiang-Nan Jin$^1$\footnote{
xnjin@ucas.ac.cn	
} and  Chia-Wei Liu$^{1,2} $}
\affiliation{$^1$School of Fundamental Physics and Mathematical Sciences, Hangzhou Institute for Advanced Study, UCAS, Hangzhou 310024, China\\
 $^2$Tsung-Dao Lee Institute, Shanghai Jiao Tong University, Shanghai 200240, China} 

\date{\today}

\begin{abstract}
We study the direct CP asymmetries in the decays of \( B^0 \to \Xi_c^+  \overline{\Xi}_c^- \) and \( B_s^0 \to \Lambda_c^+  \overline{\Lambda}_c^- \), emphasizing the critical role of final-state interactions (FSIs). In these channels, the small energy release and annihilation topology suppress short-distance contributions while enhancing long-distance effects. By separating the decay amplitudes into S and P waves, we show that the P-wave component, carrying only a single weak phase, contributes negligibly to CP asymmetries. In contrast, the S-wave amplitude—strongly modified by FSIs—acquires a substantial strong phase that enables interference among multiple weak phases, producing large CP-violating effects. Numerically, we find sizable asymmetries of \( a_{CP}^{\text{dir}} = 0.88 \pm 0.05 \pm 0.10 \) for \( B^0 \to \Xi_c^+  \overline{\Xi}_c^- \) and \( a_{CP}^{\text{dir}} = -0.106 \pm 0.019 \pm 0.010 \) for \( B_s^0 \to \Lambda_c^+  \overline{\Lambda}_c^- \), where the first and second uncertainties stem from the poorly known hadron couplings and the omission of the \( B^0 \to K^+ K^- \to \Xi_c^+ \overline{\Xi}_c^- \) re-scattering process. 
 For practical feasibility, we restrict our analysis to intermediate states involving pseudoscalar mesons, which yield tractable loop integrals. Excited intermediate states, such as  \( D^{(*)} \overline{D}^{(*)} \), are omitted because their inclusion would require handling   non‑renormalizable interactions, introducing substantial ambiguity.  These results, combined with the relatively large branching fractions, make these modes prime targets for experimental searches at LHCb and Belle II, potentially offering new insights into strong dynamics and nonperturbative QCD in heavy-hadron decays.

\end{abstract}

\maketitle

\section{
Introduction
} 

CP violation (CPV) is a cornerstone of the Standard Model (SM) and an important aspect of searches for new physics. In the SM, CPV is closely tied to the number of fermion generations and is described by a single weak phase~\cite{Kobayashi:1973fv}. Both theoretical and experimental investigations have been carried out since the first observation of direct CP asymmetries in 2001~\cite{BCP,ParticleDataGroup:2024cfk}. Although no clear evidence has yet indicated the necessity for new physics at colliders, it is understood that a new source of CPV must exist to explain the matter--antimatter asymmetry observed in the universe~\cite{Sakharov:1967dj}.

Very recently, a $5\sigma$ evidence of CP violation in decays involving baryons has been reported~\cite{LHCb:2025ray}, with its magnitude much smaller than that in the $B$ meson system.
 On the theoretical aspect, CPV in two-body $\Lambda_b$ decays is found to be small because the tree and penguin operators contribute separately to negative and positive helicities at leading order~\cite{Liu:2021rvt}. Consequently, interference between weak phases does not manifest in the total decay branching fractions,\footnote{Equivalently, the S- and P-wave CP asymmetries can cancel each other~\cite{Han:2024kgz}.} making CPV difficult to observe. To search for CPV in baryon decays, it may therefore be advantageous to focus on systems where baryons appear in the final state instead of the initial state. In particular, charmless $B$-meson decays with two baryons in the final states have been studied~\cite{Jin:2021onb}, and sizable CPV has been predicted~\cite{Geng:2023nia,Chua:2022wmr}; however, experimentally measuring these modes is challenging because their branching fractions are very small. In this work, we propose measuring CPV in charmful two-body baryonic decays.

Final-state interactions (FSIs) are known to play a crucial role in low-energy processes, especially when the released energy is close to the QCD scale~\cite{Fajfer:2003ag,Cheng:2024hdo,Jia:2024pyb,Petrov:2024ujw}. In most $b$-hadron decays, FSIs can be safely neglected because the released energy is large. However, for processes such as $B ^0 \to \Xi_c^+\overline \Xi_c^-$, the released energy is of the same order as the QCD scale, implying that FSIs should be dominant. Notably, both $B_s^0 \to p \overline{p}$ and $D_s^+ \to p \overline{n}$ are chiral-suppressed at short distances and are predicted to have branching fractions around $10^{-10}$ and $10^{-6}$, respectively~\cite{Jin:2021onb,Chua:2022wmr,Pham:1980dc,Hsiao:2014zza,Chen:2008pf,Bediaga:1991eu}. While the former is consistent with the experimental bound of $<4.4 \times 10^{-9}$, the latter disagrees significantly with the measured value of $(1.22 \pm 0.11)\times 10^{-3}$~\cite{ParticleDataGroup:2024cfk}. After incorporating FSIs, however, the prediction for $D_s^+ \to p \overline{n}$ becomes $(1.43 \pm 0.10)\times 10^{-3}$, which aligns well with the data~\cite{Geng:2024uxp}.

In this work, we investigate the impact of FSIs on the CP asymmetries of $B^0 \to \Xi_c^+  \overline{\Xi}_c^-$ and $B_s^0 \to \Lambda_c^+  \overline{\Lambda}_c^-$. 
We choose these two channels because they involve an annihilation-type topology~\cite{Rui:2024xgc}, in which short-distance contributions are suppressed.
In Sec.~II, we introduce the formalism. In Sec.~III, we present the relevant input parameters and our numerical results. Finally, we offer our conclusion in Sec.~IV.

\section{Formalism}

Our goal is to analyze the CP asymmetries of $B^0 \to \Xi_c^+  \overline{\Xi}_c^-$ and $B_s^0 \to \Lambda_c^+  \overline{\Lambda}_c^-$.
We consider these two decays exclusively, as they are dominated by annihilation-type diagrams (see Fig.~1 of Ref.~\cite{Geng:2024uxp}),  expected to be suppressed at short distances. 
 We begin with $B^0 \to \Xi_c^+  \overline{\Xi}_c^-$, while $B_s^0 \to \Lambda_c^+  \overline{\Lambda}_c^-$ is related by the U-spin symmetry. 
  The amplitude is  decomposed into two partial waves
\begin{equation}
	\langle \Xi_c^+  \overline{\Xi}_c^- | {\cal H}_{\text{eff}} | B^0 \rangle  
	= i\,\overline{u}\,\bigl(A + B \,\gamma_5 \bigr)\,v \,,
\end{equation}
where $u$ and $v$ are the Dirac spinors of $\Xi_c^+$ and $ \overline{\Xi}_c^-$, and $A$ and $B$ are related to the S- and P-wave amplitudes, respectively. 
The effective Hamiltonian is given by~\cite{Buchalla:1995vs}
\begin{equation}\label{heff}
	{\cal H}_{\text{eff}} 
	= \frac{G_F}{\sqrt{2}}
	\biggl[ 
	\sum_{q = u,c} 
	V_{qb}^*\,V_{qd} 
	\Bigl(
	C_1\,Q_1^q + C_2\,Q_2^q
	\Bigr)
	+ 
	V_{tb}^*\,V_{td}
	\sum_{i=3}^6 \,\sum_{q' = u,d,s,c} 
	C_i\,Q_i^{q'}
	\biggr],
\end{equation}
where $C_{1,2}$ are the Wilson coefficients, $V_{qq'}$ the CKM matrix elements, and
\begin{eqnarray}
	Q_1^q &=&
	\bigl(\overline{b}\,\gamma_\mu \,(1-\gamma_5)\,q \bigr)\,
	\bigl(\overline{q}\,\gamma^\mu \,(1-\gamma_5)\,d \bigr)
	\,,\nonumber\\
	Q_2^q &=&
	\bigl(\overline{q}\,\gamma_\mu \,(1-\gamma_5)\,q \bigr)\,
	\bigl(\overline{b}\,\gamma^\mu \,(1-\gamma_5)\,d \bigr)\,,
\end{eqnarray}
with summation over color implied. The explicit forms of $Q_i^{q'}$ with $i=3 \sim 6$ can be found in Ref.~\cite{Buchalla:1995vs}, but are not needed here. 
The time-independent direct CP asymmetry is defined by
\begin{equation}
	a_{CP}^{\text{dir}} = \frac{\Gamma - \overline{\Gamma}}{\Gamma + \overline{\Gamma}} \,,
\end{equation}
where $\Gamma$ is the decay width at $t=0$ and  the $\overline{\Gamma}$ denotes 
the decay width of 
the charge-conjugate process.

The S-wave amplitude is dominated by the FSIs. Short-distance suppression can be seen from the fact that, in naive factorization, the amplitude is proportional to 
\begin{equation}
	\langle \Xi_c^+  \overline{\Xi}_c^- \vert \overline{q}\,\gamma_\mu\,q \vert 0 \rangle \,
	\langle 0 \vert \overline{b}\,\gamma^\mu\,\gamma_5\,d \vert B^0 \rangle
	=
	i\,f_{B^0}\,p_B^\mu\,
	\langle \Xi_c^+  \overline{\Xi}_c^- \vert \overline{q}\,\gamma_\mu\,q \vert 0 \rangle
	= 0 \,,
\end{equation}
where $f_B$~$(p_B^\mu)$ is the $B$-meson decay constant~(momentum), and the last equality follows from the equation of motion. Since the energy release $m_{B^0} - 2m_{\Xi_c} \approx 346\,\text{MeV}$ is comparable to the QCD scale, 
the long-distance physics 
is  expected to play a very important role in this decay.  

On the other hand, the P-wave amplitude does receive contributions from naive factorization, which are proportional to $V_{cb}^*\,V_{cd}\,m_c$ (with $m_c$ the charm quark mass). Numerically, 
the $P$-wave branching fractions   are 
calculated to be~\cite{Geng:2024uxp}
\begin{equation}
{\cal B} _P ( B_s ^0 \to \Lambda_c ^+ 
\overline{\Lambda}_c ^-
) = (3.9\pm1.5) \times 10^{-6} \,,
~~~
{\cal B} _P( B  ^0 \to \Xi _c ^+ 
\overline{\Xi }_c ^-
) = (1.3\pm 0.5) \times 10^{-7} \,.
\end{equation}
Nevertheless, the P-wave amplitude is dominated by a single weak phase and therefore cannot generate CP asymmetries in the branching fractions due to lack of interference.

\begin{figure}[t]
	\begin{center}
		\includegraphics[width=0.34 \linewidth]{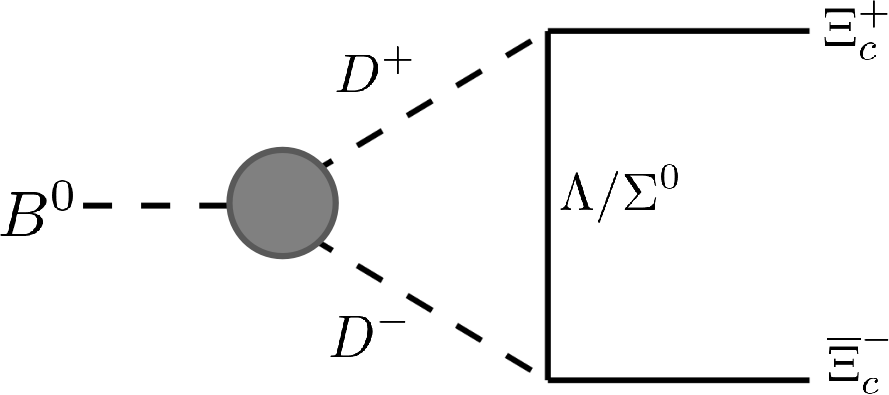}
		\includegraphics[width=0.34 \linewidth]{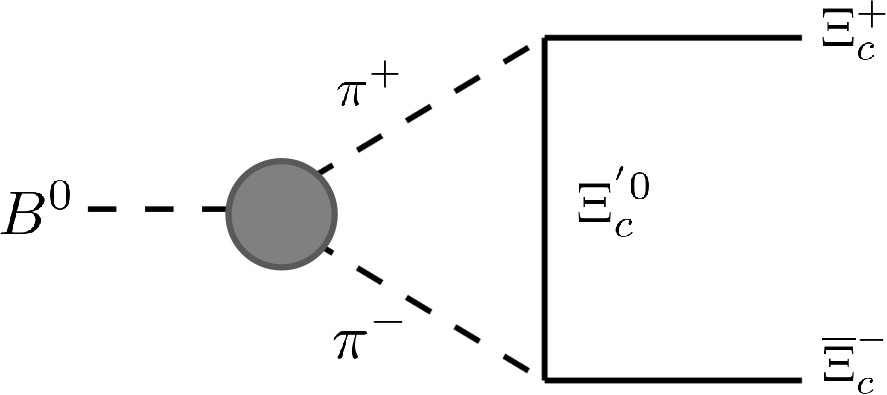}
		\caption{ 
FSI diagrams for $B^0 \to \Xi_c^+  \overline{\Xi}_c^-$. The former and latter are dominated by  the CKM elements of $V_{cb} ^* V_{cd}$ and $V_{ub}^* V_{ud}$, respectively.  
		}
		\label{fig2}
	\end{center}
\end{figure}

The considered  FSI processes  are depicted  in Fig.~\ref{fig2}. The S-wave amplitude is 
\begin{equation}\label{eq5}\small 
	A
	= 
	T_{B^0 D^+ D^-}\,
	\sum_{{\bf B} = \Lambda,\,\Sigma^0}\!
	\Bigl( 
	g_{D^+ \, {\bf B}\,\Xi_c^+}^2 \,L_c^{\bf B}
	\Bigr) 
	\,+ 
	T_{B^0 \pi^+ \pi^-}\,
	g_{\pi^+ \,\Xi_c^{\prime 0}\,\Xi_c^+}^2 \,L_u 
	\,+  
	T_{B^0 K^+ K^-}\,
	g_{K^+ \,\Sigma_c^0\,\Xi_c^+}^2 \,L_s
	\,,
\end{equation}
where $T_{B^0 P^+P^-} = \langle P^+P^- \vert {\cal H}_{\text{eff}} \vert B^0 \rangle$ is the transition amplitude for $B^0 \to P^+P^-$, and $g_{P{\bf B} _1 {\bf B}_2}$ denotes the relevant strong couplings defined by 
\begin{equation}
	{\cal L}_{\text{int}}
	= 
	-\,i \sum_{P,\,{\bf B}_1,\,{\bf B}_2}
	\bigl[
	g_{P{\bf B} _1 {\bf B}_2} P \bigl(\overline{{\bf B}}_1 \gamma_5{\bf B}_2\bigr)
	\bigr]
	+ \text{H.c.}. 
\end{equation}
The FSI loop integral is
\begin{eqnarray}
	L(m_1, m_2)
	&=& 
	\int \!\frac{d^4q}{(2\pi)^4}\,
	\frac{
		-\,\bigl(q - p_1\bigr)^\mu\,\gamma_\mu + m_2
	}{
		\bigl(q^2 - m_1^2\bigr)\,
		\bigl[\,(q - p_1)^2 - m_2^2\bigr]\,
		\bigl[\,(q - p_1 - p_2)^2 - m_1^2\bigr]
	}\,,
\end{eqnarray}
where $p_{1(2)}$ are the four-momenta of $\Xi_c^+ ( \overline{\Xi}_c^-)$. We have  employed  the shorthand of 
\begin{equation}
	L^B_c= L\bigl(m_{D^0},\,m_B\bigr),
	\quad
	L_u = L\bigl(m_\pi,\,m_{\Xi_c'}\bigr),
	\quad
	L_s = L\bigl(m_K,\,m_{\Sigma_c}\bigr).
\end{equation}
A cutoff of the form 
$\bigl[\,(m_2)^2 - \Lambda^2\bigr] \big/\bigl[\,(q - p_1)^2 - \Lambda^2\bigr]$ 
may be introduced into the integral $L(m_1,m_2)$ to account for the fact that hadronic descriptions break down at high energies, but we find that it only affects the numerical result at the $\sim 10\%$ level. For simplicity, we omit the cutoff in what follows.

	In this work, we focus on FSIs involving pseudoscalar meson intermediate states. This choice is guided by both theoretical control and practical feasibility. Including all possible intermediate states—such as those involving vector or scalar mesons—would significantly increase the complexity of the calculation and introduce large model-dependent uncertainties due to poorly known couplings and the lack of a renormalizable framework.
	As a representative example, one may consider FSIs involving vector mesons, {\it i.e.}, $B^0 \to D^+ D^{*-},  D^{*+} D^{*-}$, which are related to $B^0 \to D^+ D^-$ by the heavy quark symmetry. Nevertheless, due to the vector meson propagator of the form $(g^{\mu\nu} - p^\mu p^\nu / m_{D^*}^2) / (p^2 - m_{D^*}^2)$, the loop integrals diverge. More seriously, a theory with massive, non-gauged vector bosons is known to be non-renormalizable. While the aforementioned cutoff may tame the divergences, calculating everything in a consistent and unambiguous manner remains extremely challenging, if not impossible. Therefore, in this work, we omit the vector meson contributions. 

From the CKM matrix elements, one has the hierarchy 
\begin{equation}\label{hi}
	T_{B^0 D^+ D^-}
	\;\gg\;
	T_{B^0 \pi^+ \pi^-}
	\;\gg\;
	T_{B^0 K^+ K^-}.
\end{equation}
For evaluating CP asymmetries, two distinct weak phases are required, so the subleading terms must be retained. In practice, the third term in Eq.~\eqref{eq5} can still be omitted due to the hierarchy in Eq.~\eqref{hi}, as evidenced by the fact that ${\cal B}(B^0 \to \pi^+\pi^-)$ is about 100 times larger than ${\cal B}(B^0 \to K^+K^-)$.

\begin{figure}[t]
	\begin{center}
		\includegraphics[width=0.55 \linewidth]{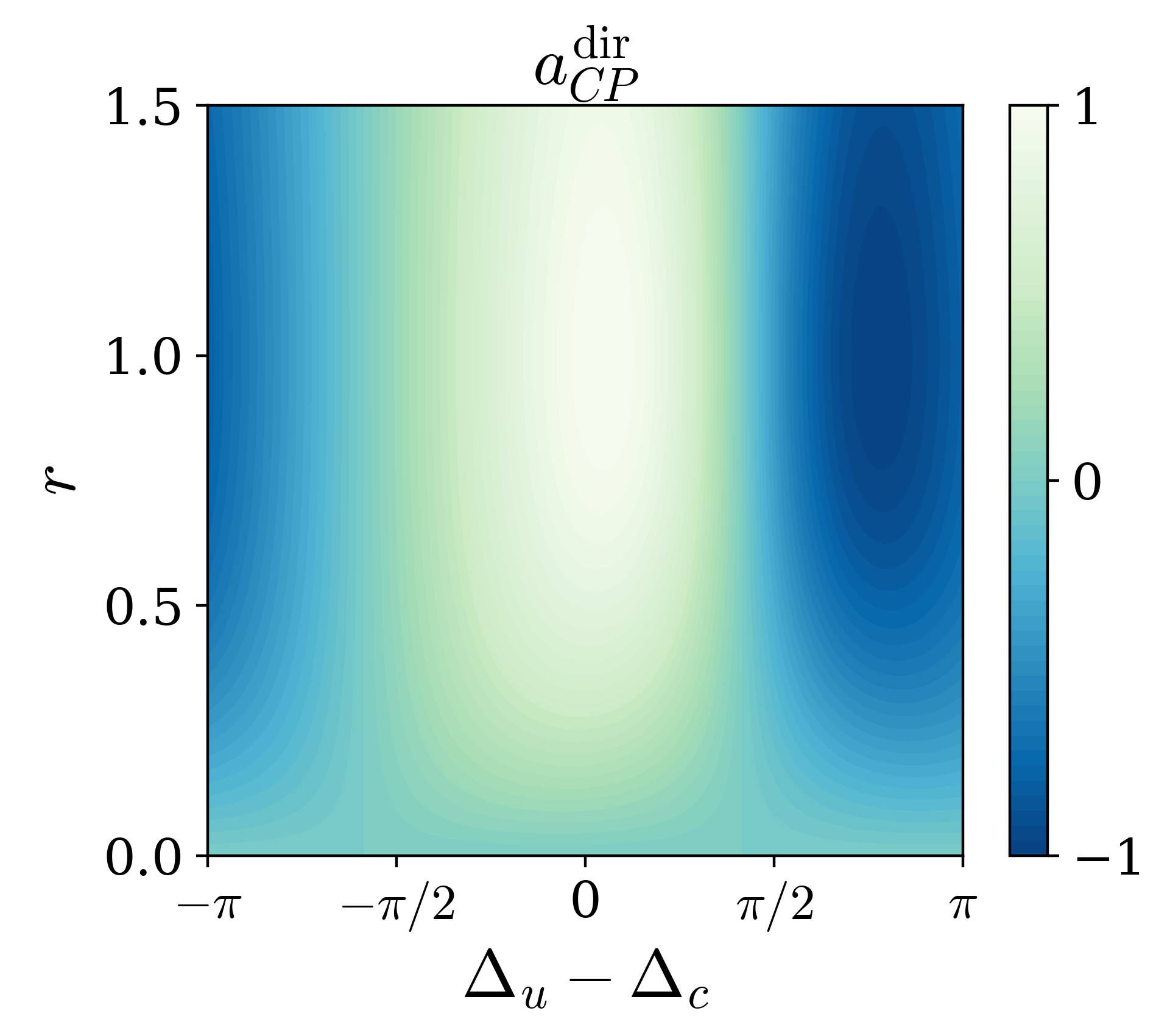}
		\caption{ 
			Dependence of $a^{\text{dir}} _{CP}$ on $r$ and $\Delta_u - \Delta_c$. 
		}
		\label{fig3}
	\end{center}
\end{figure}

Since
$|C_1 V_{cb}^* V_{cd} | > | C_1 V_{ub}^* V_{ud}| \gg | V_{tb}^* V_{td} C_{3\sim 6}| $,
the penguin contributions in
$B^0 \to D^+ D^-$ and $B^0 \to \pi^+ \pi^-$ are negligible at the level of precision considered.
The decay $B^0 \to K^+ K^-$ is primarily induced by penguin operators, and neglecting all penguin contributions collectively introduces an uncertainty of about ten percent in $\alpha_{CP}^{\text{dir}}$.
 The size of caused uncertainty is estimated by 
$\sqrt{ 
	{\cal B}(B^0 \to K^+K^-) /
	{\cal B}(B^0 \to \pi ^+\pi ^ -)  
}$.
 The short-distance  weak transitions is then  parametrized as
\begin{equation}
	T_{B^0 D^+ D^-}
	=
	e^{\,i \Delta_c}\,e^{\,i \delta_c}\,\bigl|T_{B^0 D^+ D^-}\bigr|,
	\quad
	T_{B^0 \pi^+ \pi^-}
	=
	e^{\,i \Delta_u}\,e^{\,i \delta_u}\,\bigl|T_{B^0 \pi^+ \pi^-}\bigr|,
\end{equation}
where $V_{qb}^*\,V_{qd} = e^{\,i \delta_q}\,\lvert V_{qb}^*\,V_{qd}\rvert$, and $\Delta_q$ denotes the strong phases. This leads to
\begin{equation}\label{13}
	a_{CP}^{\text{dir}}
	= 
	\frac{
		2\,r\,\sin(\delta_c - \delta_u)\,\sin    \bigl(
		\Delta _{\text{FSI}}+ 
		\Delta_u - \Delta_c \bigr)
	}{
1+ r^2 
		+ 2\,r\,\cos(\delta_c - \delta_u)\,\cos     \bigl(
		\Delta _{\text{FSI}}+ 
		\Delta_u - \Delta_c \bigr)
		+ P  _{\text{cor}} 
	}
	\,, 
\end{equation}
where $r$ is the ratio
\begin{equation}
	r \;\equiv\;
	\frac{ 
		\bigl\lvert T_{B^0 \pi^+ \pi^-}\,g_{\Xi_c'{}^{0}\,\Xi_c^+\,\pi^+}^{\,2} L_u  \bigr\rvert 
	}{
		\bigl\lvert 
		T_{B^0 D^+ D^-}\,g_{\Lambda\,\Xi_c^+\,D^+}^{\,2}L_c^\Lambda
		+ T_{B^0 D^+ D^-}\,g_{\Sigma^0\,\Xi_c^+\,D^+}^{\,2} L_c ^ \Sigma 
		\bigr\rvert
	}
	\,,
\end{equation}
$\Delta_{\text{FSI}} = \arg ( L_c^{\bf B} / L_u ) 
$  is the strong phase in FSIs,  and 
 $P _{\text{cor}} $ is a small quantity standing for the minor  correction from nonzero P-wave decay width. 
We  have utilized the $SU(3)_F$ limit of
 $ \arg (L_c^\Sigma )\approx \arg( L_c^\Lambda) $.  The parameter $r$ measures the relative strength of $B^0 \to D^+D^-$ versus $B^0 \to \pi^+\pi^-$ in the S-wave amplitude. As illustrated in FIG.~\ref{fig3}, $a_{CP}^{\text{dir}}$ vanishes if $r=0$ due to the lack of interference. Notably, $a_{CP}^{\text{dir}}$ can be nonzero even if the short-distance strong phases is absent $\Delta_u - \Delta_c = 0$, because of the presence of  
$\Delta_{\text{FSI}}$. Numerically, this phase is found to be
\begin{equation}
	\Delta_{\text{FSI}} \approx 1.84 , 
\end{equation}
which is close to \( \pi/2 \),  giving rise to a significant \( a_{CP}^{\text{dir}} \).
In fact, \( \alpha_{CP}^{\text{dir}} \) reaches its maximum at \( \Delta_u - \Delta_c = 0 \), as \( \Delta_{\text{FSI}} \) already provides the almost largest possible CP-even phase.
It is important to note that \( \Delta_{\text{FSI}} \) is independent of the input for the hadron coupling, demonstrating the capability of FSIs to induce sizable CP asymmetries.

For \(B_s^0 \to \Lambda_c^+ \overline \Lambda_c^-\), the U-spin symmetry, which describes the symmetry between the \(d\) and \(s\) quarks, gives us the relation
\begin{eqnarray}  
	\alpha_{CP}^{\text{dir}}  ( B_s^0 \to \Lambda_c^+ \overline{\Lambda}_c^- ) 
&	=&  - 
{\cal R}
	\alpha_{CP} ^{\text{dir}}  ( B^0 \to \Xi_c^+ \overline{\Xi}_c^- ) 
\nonumber\\
{\cal R} &=&
	\frac{
		|A_ {B ^0} 
		|^2
		+|  A_ {\overline B ^0}|^2
		+ 2 \kappa ^2  |   B_{B ^0 }   |^2
	}{
	|A_ {B_s^0} 
	|^2
	+|  A_ {\overline B_s^0}|^2
	+ 2 \kappa ^2  |   B_{B_s ^ 0}   |^2 }
	\,,
\end{eqnarray}
where \( A_{B_{(s)}^0} \) 
and 
\( A_{\overline B_{(s)}^0} \) 
represent  the decay  amplitudes
of $B^0 \to \Xi_c^+ \overline{\Xi}_c^-$ $(B_s^0 \to \Lambda_c^+ \overline{\Lambda}_c^-)$ and its charge conjugate process, respectively. 
Here $B_{B_{(s)}} ^0$ is the P-wave amplitude and $\kappa
 = m_B / \sqrt{
m_B^2 - 4 m  _{\bf B}^2  
}
$ 
with $m_B~(m_{\bf B})$ the mass of  the initial    meson (final   baryon). We have taken that CP is conserved in P waves. 
The above equation  is derived from the  U-spin equality of
\begin{eqnarray}
	\langle  \Xi_c^+ \overline { \Xi}_c^- | 
	(\overline{b} q ) (\overline q d)  | B ^0 \rangle 
	&=& 
	\langle  \Lambda_c^+ \overline { \Lambda }_c^- | 
	(\overline{b} q ) (\overline q s)  | B _s ^0 \rangle  \nonumber\\ 
	\sum_{q'}^{u, d, s}
	\langle  \Xi_c^+ \overline { \Xi}_c^- | 
	(\overline{b} q' ) (\overline q ' d)  | B ^0 \rangle 
	&=& 
	\sum_{q'}^{u,d,s}
	\langle  \Lambda_c^+ \overline { \Lambda }_c^- | 
	(\overline{b} q ' ) (\overline q '  s)  | B _s ^0 \rangle
\end{eqnarray}
for \( q = u,c \) and arbitrary Lorentz structure.
Hence, it suffices to calculate the CP-even quantity of 
${\cal R}$. 
From data, 
$
{\cal B} (B_s^0 \to D_s^+ D_s^-) 
= (4.4\pm 0.5) \times 10 ^{-3} 
$
is about two  hundred times larger than 
$
{\cal B} (B_s^0 \to KK)
= (2.72 \pm 0.23 ) \times 10 ^{-5} 
$
and ten thousand times larger than
${\cal B} (B_s^0 \to \pi \pi) 
= ( 7.2 \pm 1.0 ) \times 10 ^{-7} 
$~\cite{ParticleDataGroup:2024cfk},  so we consider only the intermediate state of \( B_s^0 \to D_s^+ D_s^- \)
for calculating ${\cal R}  $.

\section{Numerical results}

For the weak transitions $B\to PP$, we use the following experimental data to determine $T_{B\,P^+P^-}$:
\begin{eqnarray}
	{\cal B}(B^0 \to D^+D^-) &=& (2.11 \pm 0.18)\times 10^{-4}, 
	\quad 
	{\cal B}(B^0 \to \pi^+\pi^-) = (5.37 \pm 0.20)\times 10^{-6},
	\nonumber\\
	{\cal B}(B_s^0 \to D_s^+D_s^-) &=& (4.4 \pm 0.5)\times 10^{-3}.
\end{eqnarray}
The couplings of the $D$ meson to baryons are taken from the light-cone sum rule~\cite{Khodjamirian:2011sp}:
\begin{equation}\label{un}
	\sqrt{\frac{3}{2}}\;g_{D_s^+\Lambda_c^+\Lambda}
	= 
	-\,\sqrt{2}\;g_{D^+\Xi_c^+\Sigma^0}
	= 
	\sqrt{6}\;g_{D^+\Xi_c^+\Lambda}
	= 
	g_{D^+\Lambda_c^+n}
	= 
	10.7^{+5.3}_{-4.3},
\end{equation}
where the first three equalities follow from $SU(3)_F$ symmetry. For the pion--baryon coupling, we use the Goldberger--Treiman relation
\begin{equation}
	g_{P {\bf B}_1 {\bf B}_2} 
	= 
	G^P_{{\bf B}_1 {\bf B}_2} \frac{1}{f_P}\,\bigl(m_{{\bf B}_1} + m_{{\bf B}_2}\bigr),
\end{equation}
where $G^P_{{\bf B}_1 {\bf B}_2}$ is the form factor and $f_P$ is the meson decay constant:
\begin{equation}
	\langle {\bf B}_2 \vert \overline{q'}\,\gamma_\mu\,\gamma_5\,q \vert {\bf B}_1 \rangle 
	=
	\overline{u}_2 \Bigl( 
	G^P_{{\bf B}_1 {\bf B}_2}\,\gamma_\mu + H_T\,\sigma_{\mu\nu}(p_1-p_2)^\nu 
	+ g_3\,(p_1-p_2)_\mu 
	\Bigr)\gamma_5\,u_1,
\end{equation}
and
\begin{equation}
	i\,f_P \,q^\mu
	= \langle P \vert \overline{q'}\,\gamma_\mu\,\gamma_5\,q \vert 0 \rangle.
\end{equation}
Here, $u_{1,2}$ and $p_{1,2}$ are the Dirac spinors and momenta of $B_{1,2}$, respectively, and $q^\mu$ is the momentum of the pseudoscalar $P$.

We now discuss the possible intermediate charmed baryons in the re-scattering process of $B^0 \to \pi^+ \pi^- \to \Xi_c^+ \overline{\Xi}_c^-$.
In the heavy-quark limit, the diquarks in $\Xi_c^0$ and $\Lambda_c^+$ have $J^P = 0^-$, and the charm quark does not participate in the form factor of  $G_{\Xi_c \Xi_c}^\pi$. Hence, $\Xi_c^0 \to \Xi_c^+ \pi^-$ corresponds to $0^+ \to 0^+\,0^-$, which violates parity~\cite{Yan:1992gz}. Consequently, $G_{\Xi_c \Xi_c}^\pi = 0$. A similar argument applies to $G_{\Lambda_c^+ \Xi_c^0}^{K^+} = 0$ for $\Lambda_c^+ \to \Xi_c^0 K^+$.
On the other hand, the diquarks in the sextet baryons of $\Xi_c'$ and $\Sigma_c'$ have spin~1, and are not subject to the same heavy-quark-limit constraint. For the heavy-flavor-conserving form factors, we adopt the homogeneous bag model results~\cite{Cheng:2022jbr} 
\begin{equation}
	G_{\Xi_c'{}^+\,\Xi_c^0}^{\pi^-} = -\,0.436,
	\quad
	G_{\Sigma_c^0\,\Lambda_c^+}^{\pi^-} = 0.614.
\end{equation}
The uncertainties in the model parameters (around 10\%) are neglected here.

Notice that $B^0 \to D^+ D^-$ and $B^0 \to \pi^+ \pi^-$ share the same tree and $W$-exchange topologies. We therefore take $\Delta_u - \Delta_c = 0$ as a preliminary estimate.
The calculated branching fractions are given by
\begin{equation}
	{\cal B} (B^0 \to \Xi_c^+ \overline{\Xi}_c^-)
	= (5.6_{-3.8}^{+17.6}) \times 10^{-6} , \quad
	{\cal B} (B_s^0 \to \Lambda_c^+ \overline{\Lambda}_c^-)
	= (2.9_{-2.5}^{+11.7}) \times 10^{-4} .
\end{equation}
The uncertainties mainly arise from $g_{D{\bf B}c {\bf B}}$ in Eq.~\eqref{un}, as the amplitudes depend quadratically on this coupling. To reduce the uncertainties, we may consider the upper bound ${\cal B}(B_s^0 \to \Lambda_c^+ \overline{\Lambda}_c^-) < 8 \times 10^{-5}$, which corresponds to $g_{D^+ \Lambda_c^+ n} < 7.5$ in our framework. In the following, we will calculate the CP asymmetries, taking $6.4 < g_{D^+ \Lambda_c^+ n} < 7.5$, where the lower bound comes from light-cone sum rules.

Using the parameters above, we obtain
\begin{equation}
	r = 1.35 \pm 0.21\, \quad {\cal R} = 0.12 \pm 0.02\,.
\end{equation}
It is interesting to observe that $r > 1$, indicating that the intermediate $\pi^+ \pi^-$ state plays a more important role than $D^+ D^-$ due to the strong coupling between pions and baryons.
However, this is not the case for $B_s^0 \to \Lambda_c^+ \overline{\Lambda}_c^-$ due to the hierarchy of CKM matrix elements.
The small value of ${\cal R}$ suggests $\alpha_{CP}^{\text{dir}}$ of $B_s^0 \to \Lambda_c^+ \overline{\Lambda}_c^-$ is generally one order of magnitude smaller.
We find a large CP asymmetry:
\begin{eqnarray} 
	a_{CP}^{\text{dir}}\bigl(B^0 \to \Xi_c^+  \overline{\Xi}_ c^-\bigr)
	&=& 
	0.88 \pm 0.05 \pm 0.10\,, \nonumber\\
	a_{CP}^{\text{dir}}\bigl(B_s^0 \to \Lambda_c^+  \overline{\Lambda}_c^-\bigr)
	&=&
	-0.106 \pm 0.019 \pm 0.010\,.
\end{eqnarray}
The key to this sizable CP asymmetry lies in $r\approx 1 $ and the substantial strong phase from FSIs, which together lead to significant interferences.
The first uncertainty arises from the poorly known coupling $g_{D^+ \Lambda_c^+ n}$, while the second comes from neglecting the re-scattering process $B^0 \to K^+ K^- \to \Xi_c^+ \overline{\Xi}_c^-$.
Since $B^0 \to \Xi_c^+ \overline{\Xi}_c^-$ and $B_s^0 \to \Lambda_c^+ \overline{\Lambda}_c^-$  have sizable branching fractions, they can be experimentally probed at both LHCb and Belle~II, providing a promising opportunity to test the significance of FSI-induced strong phases and to search for sizable CP-violating effects.

\section{Conclusion}
We have explored the direct CP asymmetries in the decays of \( B^0 \to \Xi_c^+  \overline{\Xi}_c^- \) and \( B_s^0 \to \Lambda_c^+  \overline{\Lambda}_c^- \) by taking into account the crucial role of FSIs. Due to the small energy release and the corresponding QCD scale, short-distance contributions are suppressed, whereas long-distance effects become significant. The P-wave amplitude, although receiving contributions from naive factorization, does not lead to sizable CP asymmetries because it carries only a single weak phase. In contrast, the S-wave amplitude---intensely modified by FSIs---acquires a large strong phase, which allows the interference of different weak phases and ultimately gives rise to large CP-violating effects. For theoretical control and practical feasibility, we have restricted our analysis to intermediate states involving pseudoscalar mesons, which have led to tractable loop integrals. Excited intermediate states such as \( D^{(*)} \overline{D}^{(*)} \) have been omitted, as their inclusion would have required dealing with divergent loop integrals and non-renormalizable interactions, introducing substantial ambiguity.

Our numerical analysis indicates sizable direct CP asymmetries of 
\( a_{CP}^{\text{dir}} = 0.88 \pm 0.05 \pm 0.10 \) for 
\( B^0 \to \Xi_c^+ \overline{\Xi}_c^- \) and 
\( a_{CP}^{\text{dir}} = -0.106 \pm 0.019 \pm 0.010 \) for 
\( B_s^0 \to \Lambda_c^+ \overline{\Lambda}_c^- \), 
with the first and second uncertainties arising from the hadronic coupling and the omission of the re-scattering process \( B^0 \to K^+ K^- \to \Xi_c^+ \overline{\Xi}_c^- \), respectively. 
Crucially, the generation of a significant strong phase through FSIs—comparable to or exceeding \( \pi/2 \)—is what renders such large CP asymmetries feasible.
Given their relatively large branching fractions, these channels offer promising opportunities for experimental verification at LHCb and Belle II, where precise measurements of CP asymmetries would provide deeper insights into nonperturbative QCD effects and the intricate role of strong dynamics in heavy-hadron decays.

\begin{acknowledgements}
This work is supported in part by the National Key Research and Development Program of China under Grant No. 2020YFC2201501 and  the National Natural Science Foundation of China (NSFC) under Grant No. 12347103 and 12205063.
\end{acknowledgements}

\end{document}